\documentclass[fleqn,10pt]{wlscirep_arxiv}

\title{Fracture of a model cohesive granular material}

\author[1]{Alexander Schmeink}
\author[1,2]{ Lucas Goehring}
\author[1,3]{Arnaud Hemmerle}
\affil[1]{Max Planck Institute for Dynamics and Self-Organization (MPIDS), G\"ottingen, 37077, Germany}
\affil[2]{School of Science and Technology, Nottingham Trent University, Clifton Lane, Nottingham, NG11 8NS, UK; E-mail: lucas.goehring@ntu.ac.uk.}
\affil[3]{Present address: 
Aix-Marseille Universit{\'e}, CNRS UMR 7325 (Centre Interdisciplinaire de Nanosciences de Marseille - CINaM), Marseille Cedex 9, France; E-mail: hemmerle@cinam.univ-mrs.fr.}

\begin{abstract}
We study experimentally the fracture mechanisms of a model cohesive granular medium consisting of glass beads held together by solidified polymer bridges. The elastic response of this material can be controlled by changing the cross-linking of the polymer phase, for example. Here we show that its fracture toughness can be tuned over an order of magnitude by adjusting the stiffness and size of the polymer bridges. We extract a well-defined fracture energy from fracture testing under a range of material preparations.  This energy is found to scale linearly with the cross-sectional area of the bridges. Finally, X-ray microcomputed tomography shows that crack propagation is driven by adhesive failure of about one polymer bridge per bead located at the interface, along with microcracks in the vicinity of the failure plane. Our findings provide insight to the fracture mechanisms of this model material, and the mechanical properties of disordered cohesive granular media in general.
\end{abstract}
\begin{document}

\flushbottom
\maketitle

\section{Introduction}
\label{sec:intro}
Controlling the mechanical and fracture properties of cohesive porous materials is of interest for applications in various fields of research and industry, such as rock mechanics,\cite{Dvorkin1991,Jaeger2009,Holtzman2012a} the fracture of concrete \cite{Bazant1983,Buyukozturk1998} and biomaterials,\cite{Topin2008} soil rheology,\cite{Mitchell1976} geoengineering,\cite{Olson2009,Turcotte2014} or powder aggregation.\cite{Rumpf1962,Pietsch1969,Kendall2001}
Sandstone is a porous rock composed of cemented grains of sand, for example, and its fracture, induced by physical, chemical, or biological processes,\cite{Bemand2013,Warscheid2000} is highly undesirable when used in man-made structures and sculptures. On the other hand, hydraulic fracturing of underground gas-saturated sandstone is used to enhance the recovery of natural gas,\cite{Turcotte2014} for instance, in which case it is not only necessary to induce fracture propagation within the porous material, but also of primary importance to control it to limit harmful environmental side-effects. \\
\indent As a class of materials cohesive granular media can be soft, like wet sand, or extremely stiff, as porous rocks or sintered glass often are. A full understanding of the fracture properties of a cohesive porous medium requires a link between its macroscopic properties, such as its elastic moduli, yield stress and fracture toughness, and a microscopic description of its constituents, including information on the spatial distribution of its grains, and the strength of the bonds linking them.
 Crack propagation in cohesive granular media is a complex problem, given all the possible routes for breaking a cemented aggregate \cite{Affes2012,Langlois2014,Potyondy2004}  (\textit{e.g.} grains crushing, cement breaking, or bonds fully or partially detaching) and the discrete nature of granular materials leading to force chains,\cite{Cates1998} strain localization,\cite{Finno1997} or catastrophic failure,\cite{Biswas2016} for example.
 While corresponding experiments on plasticity and fracture of model systems exist for some cohesive\cite{Delenne2004,Delenne2009,Delenne2011,Langlois2014} or non-cohesive\cite{Lebouil2014,MacMinn2015} granular materials, they are still rare. However, bottom-up approaches on simple systems are highly desirable to develop constitutive laws, test numerical models, and study independently the various processes involved in fracture of granular media.\\
 \indent We have recently developed such a model system for investigating the mechanical properties of cohesive granular materials.\cite{Hemmerle2016b} The material consists of glass beads held together by rigid bridges of a solidified elastomer, whose Young's modulus, $E_p$, can be easily varied by changing its composition. The high tunability of this material allows us to focus on the contribution of the inter-particle bonds on the elasticity of the composite system, as the stiffness, strength, size, and spatial distribution of these bonds can be modified over wide ranges. It is also intended as a model system for investigations on, for example, hydraulic fracturing and biological weathering of porous media, for which a precise knowledge of both how a complex material breaks, and its resistance to fracture, are required. Furthermore, the structure of this material, which essentially consists of hard spheres connected by tunable springs, shares similarities with other disordered cohesive granular media where 
cohesion is also ensured by soft interactions between solid particles, such as charged\cite{Chen2016} or wet\cite{Herminghaus2005} granulates,
powder aggregates\cite{Rumpf1962,Kendall2001} and green bodies,\cite{Rahaman2007} and, to some extent, colloids.\cite{Russel1989, Goehring2013} It can thus provide a toy model for investigations of the failure properties of this broad class of soft materials.\\
\indent In the present work we will show how the fracture toughness of this model cohesive granular material can be varied over about an order of magnitude by changing the stiffness and volume fraction of its polymer bridges. We also extract a well-defined fracture energy of about 10 J/$\rm m^2$ from fracture testing under a range of material preparations, and demonstrate with simple scaling arguments that this energy is proportional to the cross-sectional area of the  bridges broken during crack growth. Using X-ray microcomputed tomography, we explore further the links between the macroscopic behavior of the material and its microscopic properties.  We show that cracks propagate by debonding of only about one bridge, or contact, per bead at the crack interface, and quantify how much damage occurs within the solid.

\section{Materials and Methods}
\label{sec:Methods}
\subsection{Cohesive granular material}
\label{sec:CGM}
\begin{figure}[h!]
\centering
  \includegraphics[width=90mm]{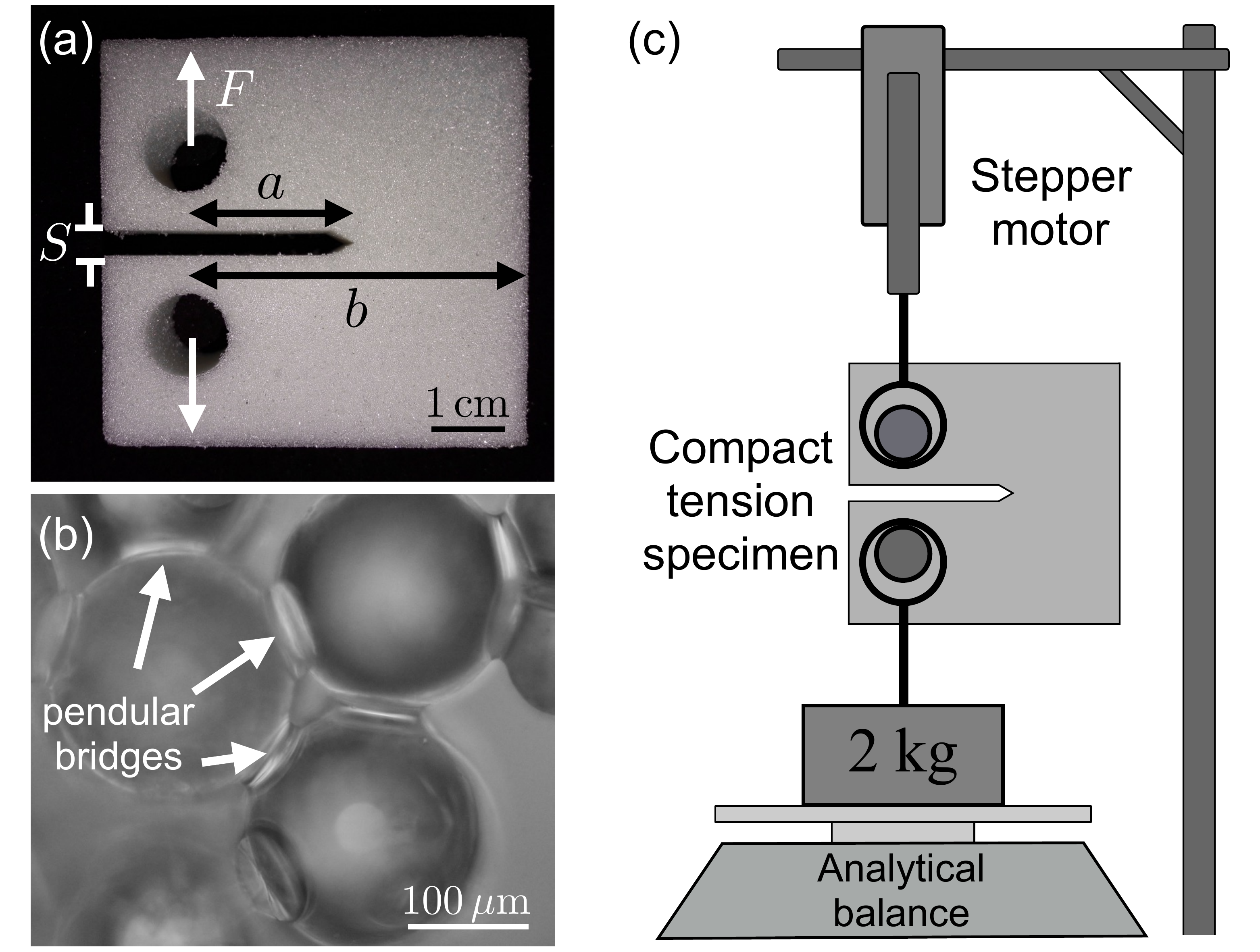}
  \caption{(a) Compact tension test specimen with cylindrical holes and pre-crack notch. When applying a force $F$ to the bars, stress is focused onto the tip of the notch.  (b) In the pendular regime of liquid content, capillary bridges of PDMS are formed between the randomly packed beads.\cite{Hemmerle2016b} The curing process solidifies these bridges, yielding a cohesive granular material. (c) Sketch of the experimental setup for the compact tension tests. While the stepper motor is moving upwards, the applied force is measured by the analytical balance and a digital camera is filming the notch opening.}
\label{im:specimen}
\end{figure}

We briefly summarize here the preparation and the main properties of the model cohesive granular material, which were detailed in a previous study.\cite{Hemmerle2016b} We use glass beads made of soda-lime glass (Sigmund Lindner) with a Young's modulus of 60-70 GPa (as stated by the manufacturer). We measure the density of the beads to be $2495\pm5\, \rm kg/m^3$ and the bead diameter to be $210 \pm 11 \mu \rm m$. The beads are thoroughly cleaned and dried following the protocol in Hemmerle \textit{et al.}~\cite{Hemmerle2016b} The dried glass beads are then mixed with polydimethylsiloxane (PDMS, Sylgard 184, Dow Corning), a cross-linkable polymer. The liquid polymer perfectly wets the beads, and forms capillary bridges between them. This process yields a homogeneous paste, which can be moulded into any desired shape.  Cross-linking of the polymer phase is then performed by baking the samples directly in a mold at 75$^\circ$C for 14 hours, and results in a solid cohesive porous material (see Fig.\,\ref{im:specimen}a,b).\\
\indent As sold PDMS consists of two parts: a viscous liquid or base, and a cross-linker.  By varying the mass ratio of the base to the cross-linker, between 50:1 and 10:1, the Young's modulus of PDMS, $E_p$, can be controlled from about $20$ kPa to $1.5$ MPa.\cite{Hemmerle2016b,Nase2013,Ochsner2007} The mechanical properties of the composite material are largely set by the stiffness of its polymer bridges,\cite{Hemmerle2016b} and its Young's modulus $E$ can be tuned from about $200$ kPa to $10\,$MPa, by varying $E_p$ between its two limits.\\
\indent The bulk material's stiffness can also be controlled by changing the PDMS volume fraction $W$. Above $W=0.5\%$ capillary bridges between beads are well defined (see Fig.\,\ref{im:specimen}b) and grow in size with increasing $W$, until they start to merge into clusters\cite{Scheel2008,Herminghaus2005,Hemmerle2016b} at the so-called pendular-funicular transition, namely around $W\simeq 3\%$. The results presented here all belong to the pendular regime, between these two limits. It has been shown previously that the average cross-sectional area of the bridges, $A$, and the Young's modulus of the material, $E$, both scale linearly with $W$ within the pendular regime.\cite{Hemmerle2016b} \\
\indent Finally, the packing fraction of the beads was measured in each cured sample tested here, and was on average $\phi_b=58.7 \pm 0.8 \%$, showing no trends with varying $E_p$ or $W$.

\subsection{Compact tension tests}
\label{sec:CT test}
\begin{figure}[h!]
\centering
\includegraphics[width=84mm]{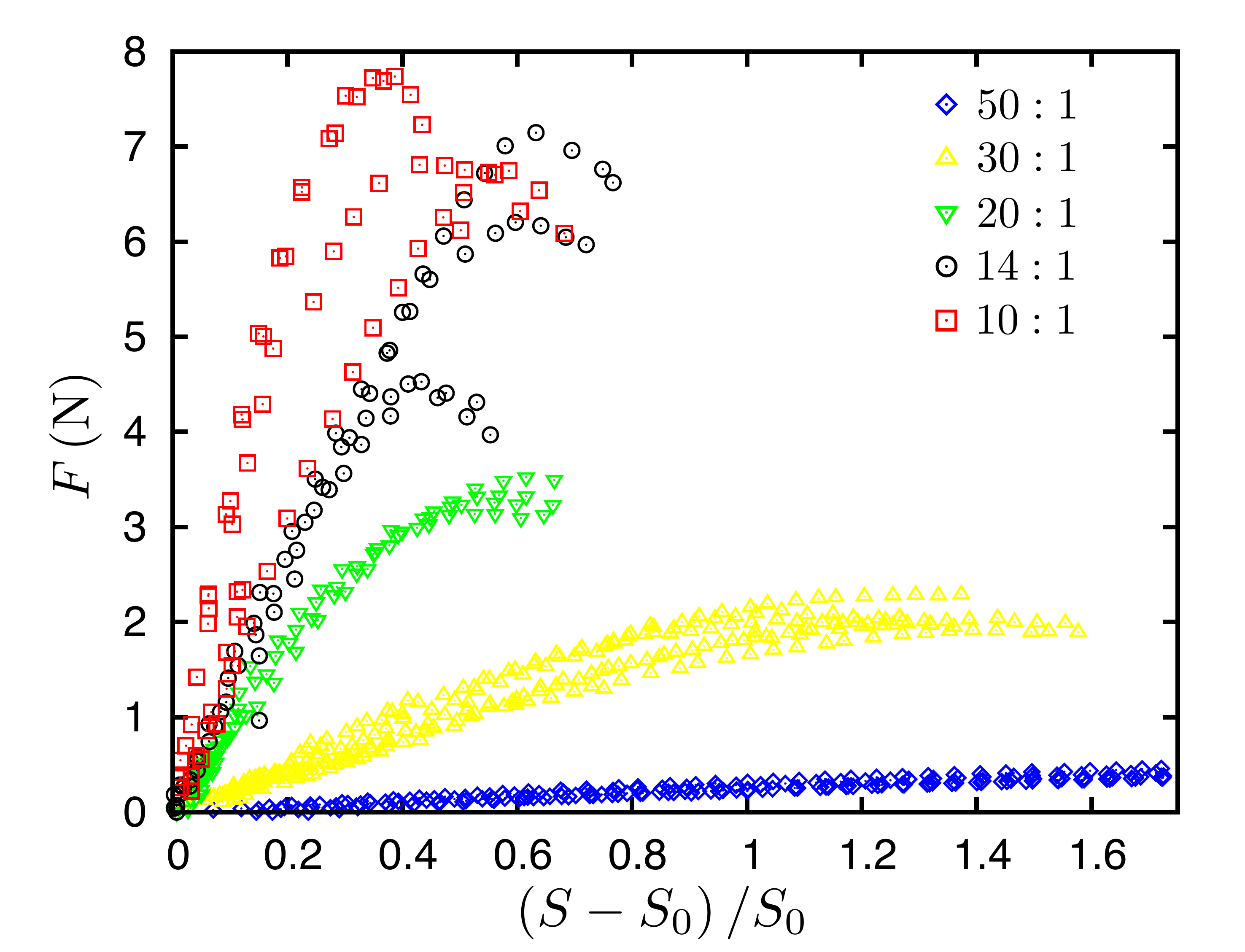}
 \caption{Force-displacement curves for several replicates at different base-to-cross-linker ratios of the polymer, from 50:1 to 10:1. $F$ is the restoring force measured on the balance, {\it i.e.}~the force applied on each bar, and $\left (S- S_0\right)/S_0$ the relative crack mouth opening displacement. Smaller mixing ratios result in stiffer polymer bridges and therefore stiffer samples and higher critical forces at failure.}
\label{im:ForceDispl}     
\end{figure}

Fracture toughness was introduced by Irwin \cite{Irwin1957} as an intrinsic material parameter quantifying the resistance of a material to the propagation of a pre-existing crack, or flaw. In the present case, we measure the Mode-I fracture toughness, $K_{Ic}$, of our material, which corresponds to the case of a crack pulled open under normal tensile stress.  To this end, we use the standard compact tension test, following the requirements of the ASTM Standard E399.\cite{ASTM2005,Tada2000} We will briefly describe here the preparation of the samples used in this fracture test, and how it allows the measurement of the Mode-I fracture toughness. \\
\indent As mentioned above, the pre-cured paste of beads and liquid PDMS is malleable and can be formed into any shape. Therefore, we baked samples directly in molds of the compact tension geometry. These test samples consist of notched rectangular blocks with two circular holes on each side of the notch, as shown in Fig.\,\ref{im:specimen}a. Two metallic bars of diameter $10$ mm are placed in the holes of the samples. The geometry of the compact tension tests ensures that when these bars are being pulled apart the stress in the sample is focused on the tip of the notch and leads to a pure Mode-I fracture opening.\cite{Lawn1993} In other words, the geometry allows for the measurement of the fracture toughness of the specimen's material, by creating a well-defined condition of tension around a crack tip. The dimensions of the specimens used here are kept fixed.  As measured from the axis of tension defined by a line joining the two bars, we use a notch of length $a$ = 20 mm and the sample extends a distance $b$ = 40 mm (see Fig.\,1a). The initial crack mouth opening length is $S_0=3$ mm and the thickness of the samples $B=20$ mm. Due to the method of moulding the specimens, $B$ may vary from its nominal value by $\pm 5\%$, and it is therefore measured and accounted for individually for each specimen. \\
\indent In our setup, the specimen rests on an analytical balance, and is held down by a 2 kg weight connected to the sample \textit{via} one bar through the bottom hole (see Fig.\,\ref{im:specimen}c). Tension was applied to the sample by a stepper motor connected to the other bar, and which moved upwards in discrete 100 $\mu$m steps. After each such step the scale was monitored by a LabView program, until the restoring force had reached an equilibrium value. Only then was the next tension step made, leading to an average notch opening speed between $2.0\;\mu$m/s and $7.7\;\mu$m/s, depending on the sample tested. These conditions ensure that the deformation is quasi-static, in which case it has been shown that the mechanical response of the material is largely independent of the strain rate.\cite{Hemmerle2016b}\\
\indent A digital SLR camera is focused on and observes the crack mouth opening, whose length $S$ is later obtained by image processing using Matlab. The applied force $F$ is measured with a resolution of $50\,$mN, while the mouth opening displacement $S-S_0$ is measured with a resolution of $50\,\mu$m. A typical force-displacement curve consists of three parts: (i) onset, where the effective strain of the mouth opening, $\delta=\left(S-S_0\right)/S_0$, is between 0 and 0.1, which corresponds to progressive contact between the bars and the sample, followed by (ii) a linear elastic response, and (iii) a plateau or peak force at failure (see Fig.\,\ref{im:ForceDispl}). Each averaged value presented here is the result of at least three replicates of such a test.\\
\indent In the compact test geometry the external stress applied to the specimen by the bars will be focused onto the tip of the notch. The divergence of stress there can be characterized by the Mode-I stress intensity factor $K_I$, which is a function of the restoring force $F$ and the geometry of the sample,
\begin{equation}
	K_I= \frac{F}{B\sqrt{b}} f\left(\frac{a}{b}\right) \label{eq:SIF},
 \end{equation}
where $f\left(a/b\right)$ is a geometric factor well-approximated by a fourth-order polynomial as specified in the literature.\cite{Srawley1976,Tada2000,ASTM2005} At some specific force, a fracture will start to grow from the tip of the notch. The critical stress intensity factor at fracture initiation, $K_{Ic}$, is the fracture toughness, and is a property that characterises a material's generic resistance to breaking.\\
\indent The distribution of stresses within the sample is highly non-uniform, due to the presence of the notch. Nevertheless, the Young's modulus of the sample can also be estimated from the linear part of the force-displacement curves \textit{via} the semi-empirical equation
\begin{equation}
 E=\frac{F}{B \left(S-S_0\right)} q\left(\frac{a}{b}\right) \label{eq:Estar}
 \end{equation}
where $q(a/b)$ is another specified geometric factor.\cite{Saxena1978,ASTM2005,Tada2000} This method for obtaining $E$ is less straightforward than in standard uniaxial tests, but both methods provide a satisfying agreement, as will be shown in Fig.\,\ref{im:KicE}a and presented in the Results section.

\subsection{X-ray microcomputed tomography}
\label{microCT}
X-ray microcomputed tomography scans (GE Nanotom) were performed on samples with a cross-sectional area of 5$\times$5 mm$^2$ and thickness of $4-5$ mm, cut from one side of the fractured interface of two specimens after completion of their fracture tests.  Tomograms were acquired with a tungsten target and acceleration voltage of 120 keV. A scan consists of a set of 3400 projections with a resolution of 2288$\times$2288 pixels, and a voxel size of 1.7 $\mu$m. The high resolution of the scans allows us to clearly distinguish the PDMS bridges from the beads, and therefore to study the fracture mechanisms at the microscopic scale. In Section \ref{microfrac}, we counted manually the number of intact and broken bridges on tomograms for populations of equal numbers (117 beads), and the errors given are the standard errors of the mean. 

\section{Results \& Discussion}
\label{sec:Results}
The high tunability of the model material allows us to study how the fracture properties of cohesive granular media depend on their microscopic details. In particular, we investigated the influence of the PDMS stiffness and volume fraction on the fracture toughness of the material, $K_{Ic}$. In a first series of measurements we varied the ratio of base to cross-linker in five values between 10:1 and 50:1, while the polymer content $W$ was kept constant at $2.3\%$. Next, the Young's modulus of the polymer was kept constant at $E_p=1$ MPa, while we varied $W$ in six steps between $0.5\%$ and $2.7\%$.  Finally, we investigated the fracture mechanisms at the scale of the beads using X-ray microcomputed tomography and linked these microscopic results to the macroscopic description of failure.

\subsection{Changing polymer composition}
\label{compo}
\begin{figure}[h!]
\centering
    \includegraphics[width=70mm]{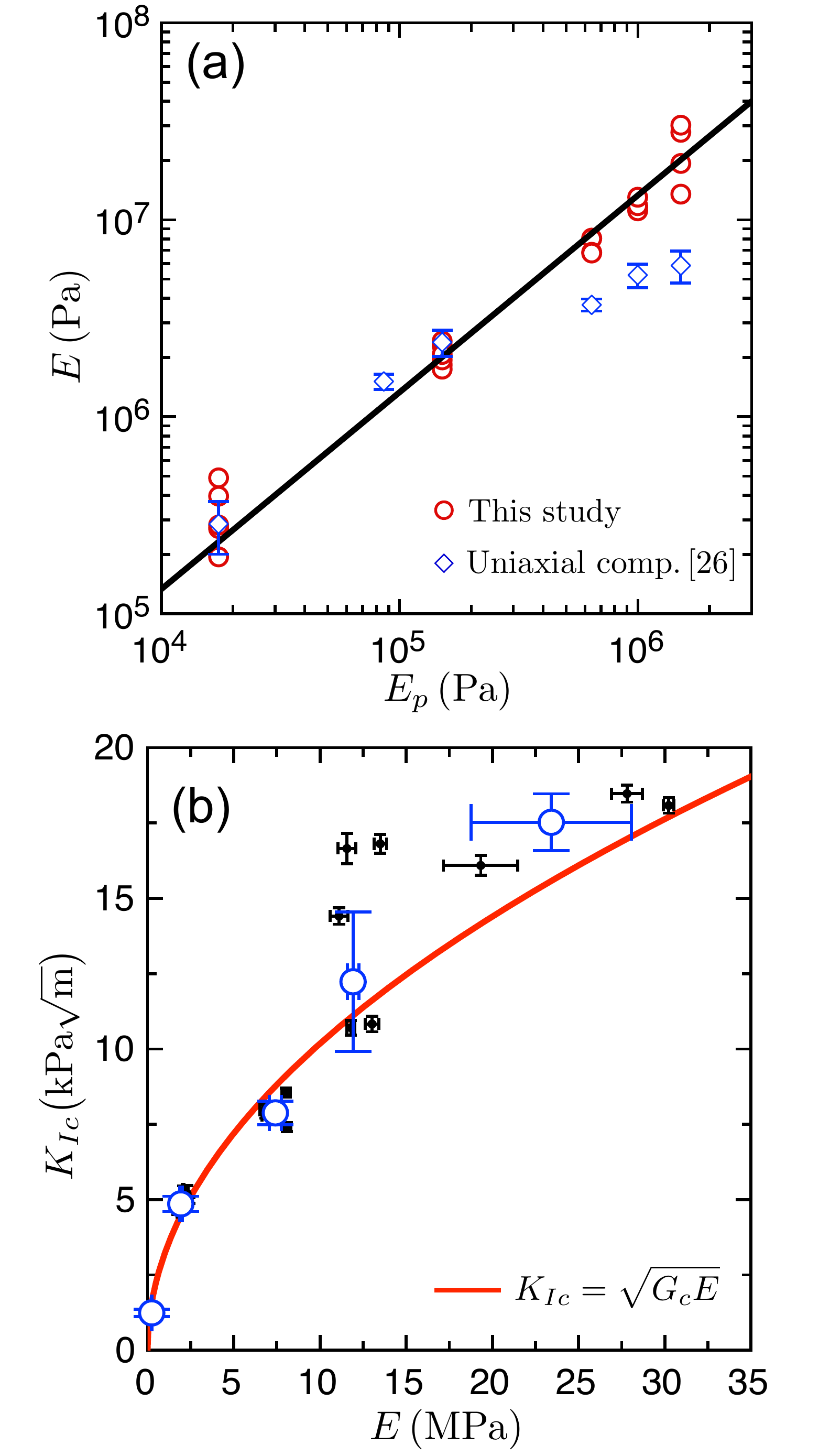}
 \caption{ (a)  The Young's modulus of the test specimens, $E$, increases along with the Young's modulus of the polymer $E_p$.  Solid line: fit of the compact tension data with a linear relation between $E$ and $E_p$. (b) The fracture toughness $K_{Ic}$ is measured as a function of the Young's modulus of the specimen $E$. Black dots show individual results for various $E$ and open blue circles are weighted means for all measurements at the same $E_p$, with standard deviations shown as error bars. Solid red line: best fit of the data using Eq.\,\ref{eq:KE}, yielding $G_c=10.4\pm0.5\, $ J/m$^2$.}
 \label{im:KicE}
\end{figure}

When varying the PDMS composition, the Young's modulus of the bridges, $E_p$, is being changed, which will consequently modify the stiffness and strength of the composite material. Here we determine how the Young's modulus $E$ and the fracture toughness $K_{Ic}$ of samples vary with $E_p$, and present this result in Fig.\,\ref{im:KicE}. We find that by changing $E_p$ from $0.02$ to $1.5$ MPa, $E$ varies across two orders of magnitude ($10^5-10^7\,$ Pa) and $K_{Ic}$ changes by about one order of magnitude ($1-18\,$ kPa$\sqrt{\rm m}$).\\
\indent In Fig.\,\ref{im:KicE}a, we compare our data for $E$ from compact tension tests with those from our previous study measuring $E$ in unconfined uniaxial compression.\cite{Hemmerle2016b} We see that the values of $E$ for both testing methods are close, except for $E_p>0.6$ MPa, for which the results in compact tension are two to four times higher than the results from uniaxial compression. Given that the inhomogeneous stress field of the compact tension test is more susceptible to non-linear effects, due to the stress divergence at the notch tip, this agreement is satisfying.\\
\indent Thus, we find that we can tune the stiffness of the material by varying $E_P$, while keeping all other parameters constant, including the statistical description of the microscopic geometry. For example, here the bead and bridge size distributions, as well as the connectivity of the bridge network, remain identical from one sample to another. In order to explain these results, we now present the measured $K_{Ic}$ as a function of $E$, and compare it to predictions of linear elastic fracture mechanics (LEFM). In LEFM, the fracture toughness $K_{Ic}$ is related to both the specimen's Young's modulus $E$ and its critical strain energy release rate $G_c$, which is the energy consumed, per unit of interfacial area, by the new surfaces generated by fracture:\cite{Irwin1957,Lawn1993}
\begin{equation}
 K_{Ic}=\sqrt{G_c E}. \label{eq:KE}
\end{equation}
Our data are in good agreement with Eq.\,\ref{eq:KE}, and the assumption that $G_c$ does not depend on $E_p$, as shown in Fig.\,\ref{im:KicE}b. This independence of $G_c$ and $E_p$ will be discussed further in Section \ref{energy}, where we compare the fracture energy with the microscopic damage that occurs along the crack surface. A fit of Eq.\,\ref{eq:KE} to our data allows us to extract a well-defined fracture energy of $G_c=10.4\pm0.5\, $ J/m$^2$ for the material at a constant $W=$2.3\%. This is of the same order of magnitude as typical fracture energies of sandstone or porcelain, for example.\cite{Kendall1987,Cotterell2010,Ashby2005}

\subsection{Changing polymer content}
\label{poly}
\begin{figure}[h!]
\centering
 \includegraphics[width=75mm]{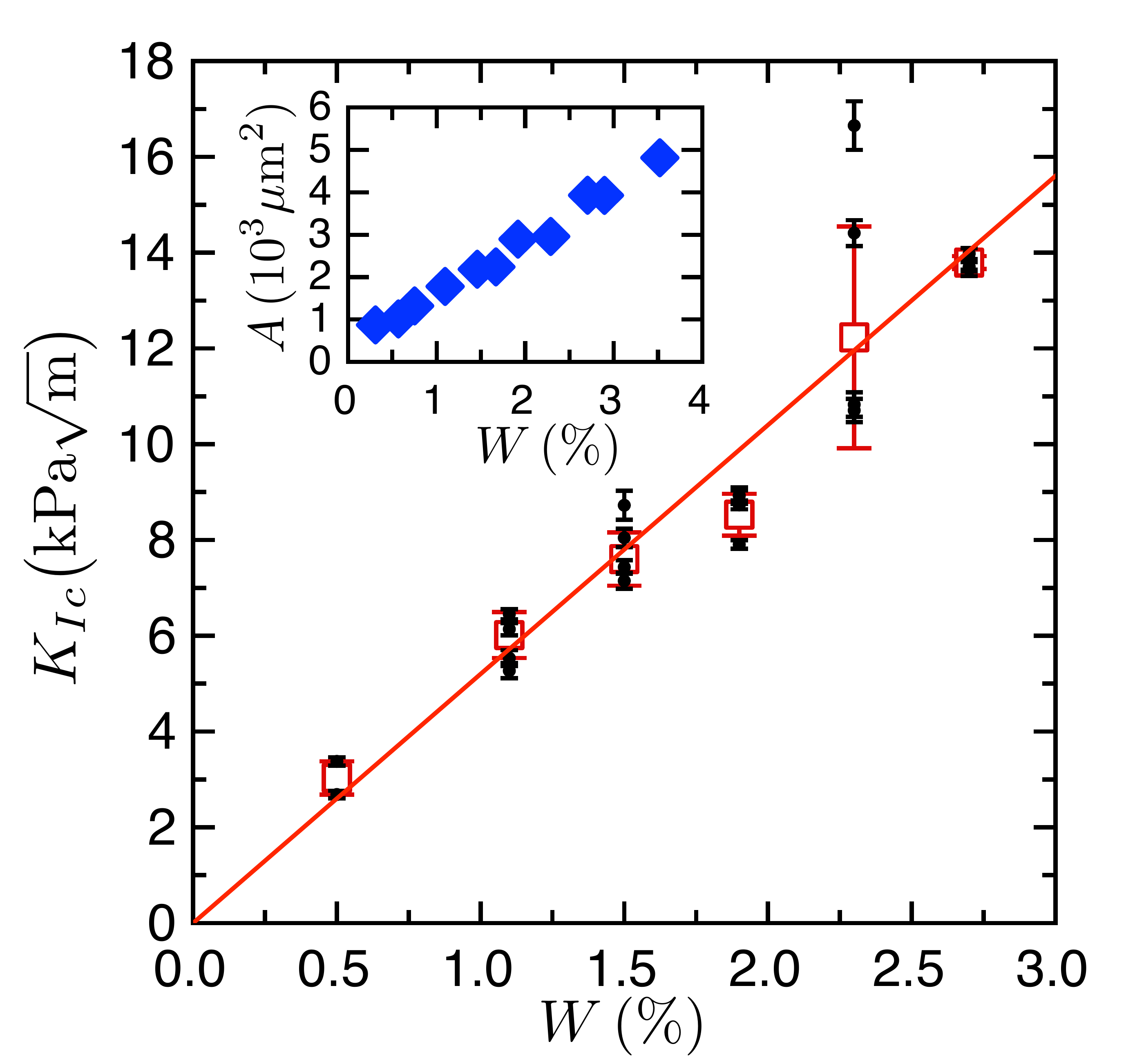}
\caption{Fracture toughness $K_{Ic}$ as a function of polymer content $W$. Black dots show individual results and open red squares show the weighted means calculated for all $K_{Ic}$ that share the same $W$; error bars give standard deviations. All $W$ lay in the pendular regime, where the liquid polymer forms isolated capillary bridges between the glass beads. We find a linear scaling relation between the two quantities, $K_{Ic} \sim W$. Inset: measurements by Hemmerle \textit{et al.}~\cite{Hemmerle2016b} show that the average bridge cross-sectional area, \textit{A}, also scales linearly with $W$.}
 \label{im:KicW}
\end{figure}

The polymer content $W$ influences the size of the capillary bridges in the material, giving us another possibility to vary the mechanical properties of the system. Tests were conducted with a constant $E_p= 1\,\rm MPa$ (a composition ratio of 14:1) and six different $W$, making sure to keep the polymer content below the pendular-funicular transition at $W\simeq 3\%$.\cite{Hemmerle2016b} By varying W between 0.5\% and 2.7\%, we can tune $K_{Ic}$ in the range of $3-14\,\rm kPa\sqrt{\rm m}$. As shown in Fig.\,\ref{im:KicW}, there is a clear linear relation between the fracture toughness and the polymer content, $K_{Ic}\sim  W$. However, it will be more practical to consider $K_{Ic}$ in relation to the area of the broken bridges, rather than their volumetric fraction. The mean cross-sectional area, $A$, of the bridges was found to grow linearly with $W$ in Hemmerle \textit{et al.}~\cite{Hemmerle2016b} (see inset in Fig.\,\ref{im:KicW}).  This implies that there is a linear relation between the fracture toughness of the material and the cross-sectional area of the bridges $K_{Ic}\sim A$.

\indent Two processes are driving the change in the fracture toughness when varying the size of the bridges. First, Hemmerle \textit{et al.}~have shown that within the pendular regime $E \sim A$, {\it i.e.\,}that the Young's modulus of the material varies linearly with the cross-sectional area of the bridges within it.\cite{Hemmerle2016b} According to Eq.\,\ref{eq:KE}, $K_{Ic}$ will thus increase with $A$, through the change in the material stiffness. Secondly, the critical strain energy release rate, $G_c$, is an energy per unit area of the cracked interface \textit{of the material} and will therefore depend on its microscopic details, such as the distribution of bridge sizes along the crack path. Larger bridges will require a higher energy to break, and consequently will lead to a higher macroscopic $G_c$. We can now estimate this relationship between $G_c$ and $A$, using our experimental measurements of scaling relations between the different parameters. 
The relations $E\sim E_p$ (Fig.\,\ref{im:KicE}a) and $E\sim A$ allow us to write $E\sim A E_p$. Assuming that, as in Section \ref{compo}, $G_c$ does not depend on $E_p$,  we can then rewrite Eq.\,\ref{eq:KE}:
\begin{equation}
 K_{Ic}\sim  \sqrt{ A E_p G_c(A)}, 
 \end{equation}
from which we deduce that, if we observe that $K_{Ic}\sim A$, then $G_c\sim A$. This result shows that the fracture energy of the cohesive granular material, measured at the sample scale, connects to the geometry of the bonds along the crack path, and more precisely to their cross-sectional area.

\subsection{Microscopic fracture mechanisms}
\label{microfrac}
\begin{figure}[h!]
\centering
  \includegraphics[width=84mm]{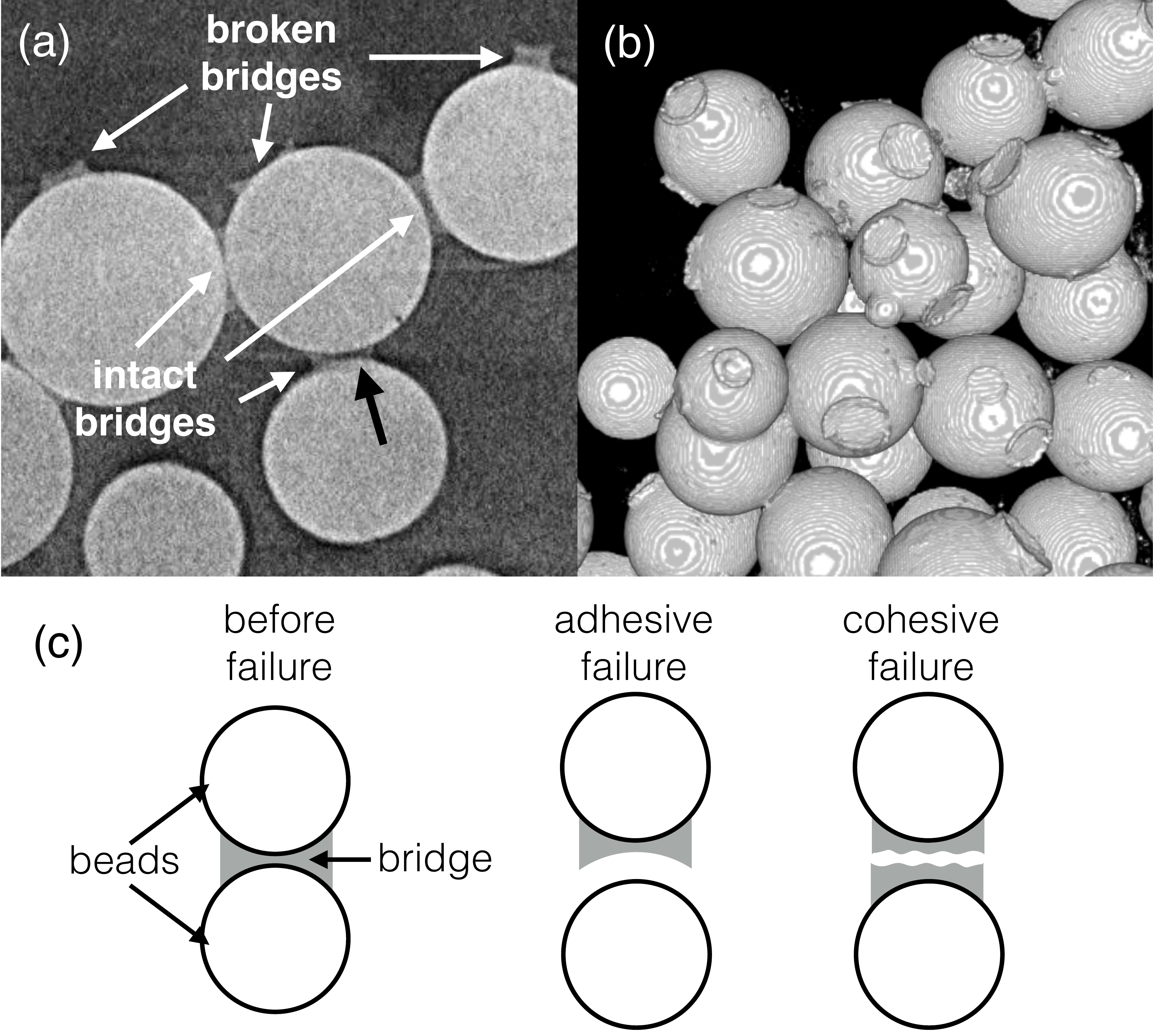}
\caption{ (a)  Fractured interface observed by X-ray $\mu$CT, here shown as a 2D slice extracted after reconstruction of the tomogram. The black arrow indicates possible partial debonding of a bridge which is still connected to two beads.  (b) 3D volume visualized after segmentation of the tomograms by manual thresholding  using ImageJ.\cite{Schneider2012} Disconnected bridges are clearly visible at the surface of the beads. (c) Illustration of the two different modes of bridge failure. }
\label{im:nanotom}       
\end{figure}

The macroscopic fracture energy of our cohesive granular medium is directly linked to the structure of its inter-particle bridges. We will now examine this correspondence from a microscopic point of view and, in particular, link the macroscopic fracture energy of the material to the energies involved in breaking the bonds between particles.  It remains to be determined, for example, if fracture is caused by the failure of the bridge material or by debonding at the bridge/particle interface, and whether all the fracture energy goes into propagating the crack plane or if it is also consumed by additional damage within the bulk of the material. To this end we used X-ray microcomputed tomograpy ($\mu$CT) to look at, and below, the fractured surfaces. The difference between the X-ray absorption coefficients of glass and of PDMS allows us to distinguish between air, beads and bridges in tomograms (see Fig.\,\ref{im:nanotom}a). \\
\indent We cut from each of two specimens a section of the fractured interface after completion of a compact tension test, and analyzed them using $\mu$CT (see Methods, section \ref{microCT}).  The samples were prepared with the two extreme polymer stiffnesses used in this study, namely $E_p=0.02$ MPa and $E_p=1.5$ MPa. Their constant $W=2.3$\% allows us to focus on any influence of the stiffness of the bridges on the fracturing process.  We see in Fig.\,\ref{im:nanotom}a that two types of bridges can be observed on the beads located at the fractured interface: bridges connected to only one bead,  further called broken bridges, and bridges still connected to two beads, further called intact bridges. The potential mechanisms of bridge failure in our system are similar to the processes involved in the failure of adhesives,\cite{Zosel1991,Galliano2003,Nase2013} and include two types of failure (see Fig.\,\ref{im:nanotom}c). One possibility is that a bridge could detach from the surface of one of the beads, leaving no trace of polymer (adhesive failure). In this case one of the two beads will lose all sign of any bridge, while the other one will keep a broken bridge. Another possibility is that fracture occurs within the bridge (cohesive failure). In this case the two affected beads will each retain a part of the broken bridge. In the following, we show that the main mechanism of fracture propagation is adhesive failure of the polymer bridges, as suggested by the curved interfaces of the broken bridges in Fig.\,\ref{im:nanotom}. We measure then the number of bridges which need to detach to propagate the crack, and finally determine how fracture changes the microstructure of the samples.\\

\indent First, we measure the average number of bridges per bead in the bulk, $N_0$, which is also known as the wet coordination number for wet granular materials.\cite{Herminghaus2013} We find that $N_0 =7.4 \pm 0.1$, with no significant difference between the softer and the stiffer samples. This is slightly higher than the typical value of 6-7 bridges per bead observed with glass beads wet by aqueous solutions at similar packing fractions,\cite{Kohonen2004,Scheel2008,Herminghaus2013} although there may be some dependence of this number on the preparation protocol.\cite{Groger2003}\\
\indent Next we count, for each bead located along the crack interface, the number of intact bridges connected to the bead, $N_i$, and the number of broken bridges present on its surface, $N_b$. We measured $N_i=5.6 \pm 0.15$ and $N_b=0.9 \pm 0.1$ on average, with again no significant difference between the two samples. For each bridge breaking \textit{via} adhesive failure, {\it i.e.\,}by debonding of the polymer from the bead surface, one of the two initially connected beads will retain the whole bridge while the other bead will lack any remains of it. Looking thus for the deficit of bridges on the interface to the bulk, we deduce that $N_0-N_i-N_b=0.9\pm 0.15$ bridges per bead broke by adhesive failure and remained attached to the bead, while the same number would have left with the opposite side of the interface. This number being equal to $N_b$, we can further conclude that essentially all the broken bridges we observe originate from debonding events, 
meaning that cohesive failure is negligible (occurring for not more than 5\% of the broken bridges, considering the error bars). \\
\indent In its present form, our experimental setup does not allow access to the dynamics of crack propagation. For example, it remains to be seen if a crack propagates as a clear front, with only the bridges at the tip debonding as the crack advances. Another possibility would be the presence of a process zone ahead of the crack, where a certain number of bridges would have to fail for the crack to extend further. We can see in the tomograms that some bridges have partially detached in the first two layers below the fracture surface.  An example of this is pointed out by the black arrow in Fig.\,\ref{im:nanotom}a. Such damage supports the hypothesis that a small process zone exists around the propagating crack front. Interestingly, the average number of intact bridges on beads along the crack path, $N_i$, is slightly below $z_c=6$, the critical coordination number in disordered 3D spring networks associated with the phase transition between floppy and rigid systems.\cite{Ulrich2013}
When approaching this transition, mechanical properties of amorphous solids changes dramatically, with, for example, a loss of stability due to a vanishing shear modulus.\cite{Ulrich2013,Driscoll2016} Fracture propagation in our material could then originate from a local loss of stability within the process zone. In other words, bridges would detach ahead of the crack tip, until the number of intact bridges in this region reaches $z_c$. This local area would then become suddenly soft and fragile, and eventually fail, propagating the crack further. 

\subsection{Fracture energy}
\label{energy}
The energy dissipated at the microscale during crack propagation must be equal to the macroscopic critical strain energy release rate measured with the fracture tests, and we will explore this equivalence here. We have seen that fracture of the material occurs almost exclusively \textit{via} detachment of the bridges from the beads surfaces, and not from internal failure of the PDMS. This result agrees well with a comparaison of the fracture energies involved in the two cases: while the (interfacial) energy required to peel PDMS off from glass is only about\cite{Chopin2011} $\Gamma_{\rm adh}\simeq 7$ J/m$^2$, the energy needed to propagate a Mode-I crack in PDMS is at least an order of magnitude higher\cite{Mills2008} $\Gamma_{\rm coh}\simeq 250 $ J/m$^2$. To compare these fracture energies to the critical strain energy release rate, $G_c$, of the composite material, we need to estimate the surface covered by the detached bridges. For both samples used in the $\mu$CT analysis we counted 
the number of beads belonging to the interface over the projected area of 2.58 mm $\times$ 2.58 mm. We found similar results for the two samples, the average being $n=22.0 \pm 0.2$ beads per  mm${^2}$ of interface. Using the mean cross-sectional area of the bridges\cite{Hemmerle2016b} $A=2960$ $\mu$m$^2$ at $W=2.3\%$, the surface of PDMS detaching per unit area of the crack is then about 12\% of the interfacial area. For a brittle crack propagating without plasticity, the macroscopic energy release rate $G_c$ would be simply the interfacial energy of the material, and originate only from the detachment of the bridges at the new interface created by the crack. We would then expect $G_c \simeq 0.12 \Gamma_{\rm adh}$, while here $G_c \simeq \Gamma_{\rm adh}$.  This again suggests that some plastic deformation occurs during fracture.  Plasticity in this case may originate from detachment of bridges in the process zone, as mentioned before, or possibly from microcracks around the main crack, as is seen the similar 
case of drying colloidal films.\cite{Goehring2013} While several broken bridges can be seen in the tomograms, it is impossible to estimate their proportion with the present experimental technique, since it is expected that most of them will recover their initial positions after deformation (but remain detached from the bead surface). \\
\indent We have seen in sections \ref{compo} and \ref{poly} that $G_c$ is independent of the stiffness of the polymer phase, but is directly proportional to the area of the polymer bridges.  This, in turn, suggests that $G_c$ depends on the geometry of the bridges only, and not on the cross-linking or compliance of the bridge material. The influence of PDMS cross-linking on its adherence has been studied previously using tack tests, by which adhesion force is measured after that a probe is put into contact to a PDMS film at a certain force and for a certain time.\cite{Galliano2003,Nase2013} Nase \textit{et al.~}have shown that for a slow probe speed of $1 \mu$m/s, the debonding energy of PDMS on steel depends only weakly on its cross-linking degree, except for full cross-linking where the adherence energy can be about 3 times smaller.\cite{Nase2013} Galliano \textit{et al.~}have found that this difference decreases significantly with the force applied prior to separation and the separation speed.\cite{Galliano2003}
However, it should be noted that the geometry of tack tests is far from our configuration, where, additionally, the PDMS is cured directly on the surface of the glass beads.\\
\indent In summary, our results suggest that the critical strain energy release rate, $G_c$, of our granular cohesive material is linked to the geometry of the bridges only, and is largely independent of the cross-linking of PDMS, and hence $E$ or $E_p$. 

\section{Conclusion}
\label{Conclusion}
We have demonstrated that the fracture toughness of a model cohesive granular material can be tuned over about an order of magnitude \textit{via} two control parameters, namely the volume fraction and Young's modulus of the bridges that bond the cohesive material together. Moreover, the repeatability of the data shows how one can aim for a specific fracture toughness of the material, and prepare it accordingly. \\
\indent By varying the elasticity of the medium, but keeping its geometry fixed, we found a well-defined critical strain energy release rate for the material, which is comparable to that of sandstone.  By then varying the polymer content of the medium, we further showed that this energy release rate scales linearly with the cross-sectional area of the polymer bridges between its grains.  The link between the macroscopic and microscopic fracture mechanisms was also investigated by X-ray microcomputed tomography, from which we have seen that the material fails \textit{via} debonding of about one bridge per bead along the crack interface. The macroscopic fracture energy is about ten times higher than the work required for this process, suggesting (along with broken bridges around the crack plane), that there is a modest amount of plastic damage during fracture. 
\\
\indent These results are important for the future use of this material to study fundamental fracture processes in granular materials, for example by hydraulic pressure, root growth, or bioweathering. More generally, the simple structure of the model material, a disordered packing of hard spheres connected by soft bonds, makes the scaling relations between its mechanical and geometrical properties, determined here, of interest for applications involving the fracturing of other disordered cohesive granular media, such as consolidated soils, colloidal materials, ceramics (green bodies) or powder aggregates.

\section{Acknowledgements}
We thank W. Keiderling for help designing the molds for the compact tension tests, and S. Biswas for fruitful discussions.

\bibliography{Schmeink_Arxiv}

\end{document}